\documentclass{optica-article}

\journal{opticajournal} 

\articletype{Research Article}

\usepackage{lineno}
\linenumbers 

\begin{document}
\nolinenumbers
\title {Enhanced Optical Vector Bottle Beams with Obscured Nodal Surfaces}
\author{Nicolas Perez\authormark{1} and Daryl Preece\authormark{1,*}}

\address{\authormark{1}University of California Irvine, Irvine, CA 92697, USA}

\email{\authormark{*}dpreece@uci.edu} 


\begin{abstract*}

Optical bottle beams, characterized by their unique three-dimensional dark core, have garnered substantial interest due to their potential applications across multiple domains of science and technology. This paper delves into the current methods used to create these beams and provides a method to obscure their nodal planes through coaxial non-interfering orthogonally polarized beams to generate bottle beams with enhanced uniformity. Experimental and theoretical results show the enhanced vector bottle beam maintains a smaller, more spherically uniform potential well and interesting quasi-particle polarization characteristics. 
\end{abstract*}

\section{Introduction}
  Optical bottle beams have emerged as a fascinating area of research, offering advantages in fields such as atom optics and quantum computing \cite{xu_trapping_2010,barredo_three-dimensional_2020}, optical trapping\cite{alpmann_holographic_2012} and microscopy\cite{shostka_optical_2019,kohl_superresolution_2013}. Optical bottle beams are comprised of a dark focus surrounded by an area of light. This counter-intuitive beam shape creates an optical potential well at the centre of the beam, (\(|E|\rightarrow 0\)). In areas such as optical tweezers, optical bottle beams are primarily used to trap absorbing particles, which are pushed out of typical Gaussian optical traps\cite{alpmann_holographic_2012}. The inverted nature of a bottle beam trap leads to reduced energy absorption by the trapped particle. This in combination with other trapping techniques has led to the trapping and cooling of atoms\cite{xu_trapping_2010}.

  First introduced by Arlt and Padgett in 1999, the concept of optical bottle beams utilized a combination of Laguerre-Gaussian modes, to create a beam with a dark central spot surrounded by a bright beam of light\cite{arlt_generation_2000}. In this regime, the radial phase discontinuity combined with the Gouy phase shift leads to destructive interference in the focal region of the beam. This early work laid the foundation for subsequent research in the field, particularly in the realm of optical trapping and atom trapping. Later, McGloin took a different approach to create the second regime by propagating higher-order Laguerre Gaussian beams through an axicon to create higher-order Bessel beams with unique bottle beam profiles \cite{mcgloin_three-dimensional_2003}. Bottle beams produced in this way can produce both single on-axis; or three-dimensional arrays of intensity minima.
  
  Great effort has also been made to improve the bottle beam symmetry and broaden their applications. Shvedov introduced the idea of generating bottle beams with white light vortices, further diversifying the field's capabilities\cite{shvedov_generation_2008}. Chremmos, by utilizing airy beams to generate larger and more homogeneous bottle beams, began addressing challenges associated with intensity variations in the focal region\cite{chremmos_fourier-space_2011}. Zhang used Moiré techniques to synthesize bottle beams from multiple vortex beams\cite{zhang_trapping_2011}. Alpmann explored the use of a Fourier convolution theorem to generate optical bottle beams with OAM\cite{alpmann_holographic_2012}. Porfirev proposed an array of optical bottle beams based on the superposition of Bessel beams\cite{porfirev_generation_2013}. Turpin explored conical refraction in biaxial crystals, allowing for the creation of bottle beams that could be easily opened and closed as needed\cite{turpin_conical_2016}. More recently, several authors have created optical bottle beams with meta-surfaces \cite{xiao_efficient_2021} and nano-jets \cite{kim_engineering_2011}. These advancements in shaping bottle beams have greatly enhanced the flexibility of the technology, especially in applications like optical trapping, atom trapping, and super-resolution microscopy, where the properties of bottle beams play a crucial role.
  
  Although the utility of these beams is obvious, most laboratory-created bottle beams are far from perfect bottles. Using interference methods to create, perfect dark regions in a beam introduces nodal surfaces. Propagation of higher-order Bessel beams also carries a similar property as the linearly propagating dark regions cross.
  
  In this paper, we investigate many of the methods used previously and utilize vectorial beam shaping techniques to improve trap uniformity, and by extension well depth, and to mitigate the discontinuities in the wave function that contribute to local intensity variations. With this in mind, we create structured beams with improved spherical uniformity and local polarization control. These improvements may lead to better trapping behaviors such as retention rate and stiffness along the nodal surfaces, and precision beam shaping for broader applications.

\section{Generation Techniques of Optical Bottle Beams}

  One of the fundamental aspects of beam shaping involves the strategic use of interference to direct light to a desired location. As waves of identical wavelength interfere they form alternating bright and dark fringes that when extended into three-dimensional space manifest as extended bright and dark surfaces. The dark surfaces, resulting from complete destructive interference, are known as nodal surfaces.
  
  Nodal surfaces then represent transform invariant regions with attenuated amplitude in wave functions which enter and exit beams continuously \cite{loiko_generating_2013,freegarde_cavity-enhanced_2002}. This is the reason bottle beams formed by interference techniques are "leaky". The dark region at the center of the beam, creating the bottle, must enter and exit the beam in some way. This leads to dark paths out of the bottle and a lobed intensity profile. Some Airy-based bottle beams can mitigate this somewhat, but significant intensity variation can still be found around the shell of the bottle\cite{chremmos_fourier-space_2011}. 

\subsection{Theory}
  While several beam formalisms exist, the beam first proposed by Arlt \cite{arlt_generation_2000} was created by the superposition of two specific Laguerre–Gaussian modes. When the relative phase of these modes is adjusted appropriately, they interfere only over the focal region, which has a non-zero Gouy phase shift, producing a beam focus with zero on-axis intensity, surrounded by regions of greater intensity. The formation of this dark focus is due to the relative phases between the combined modes, ensuring their interference leads to a reduction in intensity at the beam's center.

\begin{figure}[ht!]
  \centering\includegraphics[width=13cm]{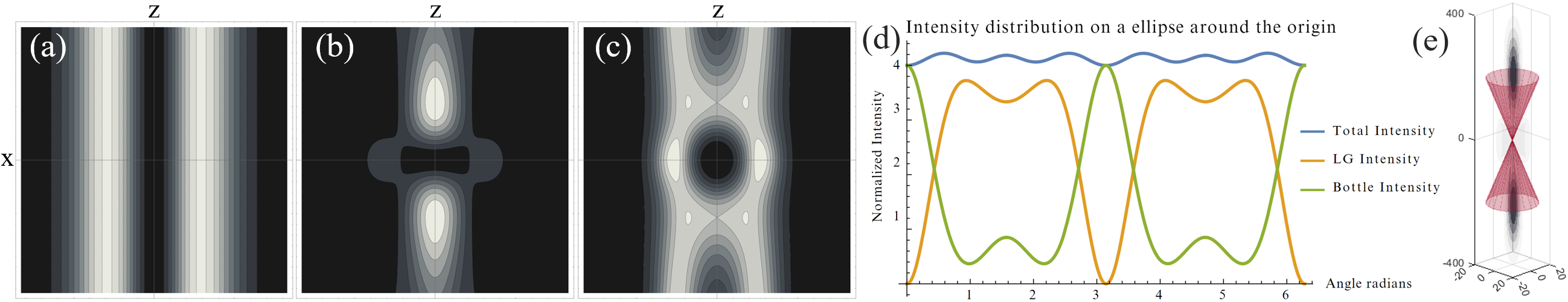}
\caption{ Contour plots of focal regions for three beams calculated using the paraxial wave equation: (a) LG beam with winding number l = 1, (b) bottle beam generated by eq. 2, (c) combination of beams (a) and (b) generated by eq. 3. (d) Elliptical intensity plot about the focal center for beams shown in (a-c). (e) Mathematical simulation of an optical bottle beam. A circular phase discontinuity, shown in Fig.3(a), is propagated by vectorial simulation through a 0.25NA lens. The projection of the nodal surface formed along the dark region following the focused phase discontinuity is shown in red.}
  \label{Fig: Thoeretical intensity distribution of LG beams}
\end{figure}
The Laguerre-Gaussian (LG) beam is typically represented as:
\begin{align}
  u_{l,p}(r, \theta, z) = C_{l,p}(w(z)) \left( \frac{r \sqrt{2}}{w(z)} \right)^{|l|} L_p^{|l|} \left( \frac{2r^2}{w^2(z)} \right) \exp \left( -\frac{r^2}{w^2(z)} \right) \exp(-il\theta) \nonumber\\ \exp \left( -i \frac{k r^2 z}{2(R(z) + i z_0^2)} \right) \exp \left( i (2p + |l| + 1) \arctan \left( \frac{z}{z_0} \right) \right)
\end{align}

Where: \( u_{l,p} \) represents the LG mode with azimuthal index \( l \) and radial index \( p \), \( C_{l,p} \) is a normalization constant, \( w(z) \) is the beam width, \( L_p^{|l|} \) is the associated Laguerre polynomial, \( R(z) \) is the radius of curvature, \( z_0 \) is the Rayleigh range, \( k \) is the wave number. Given this, a bottle beam created by interfering two LG modes \( u_{0,2} \) and \( u_{0,0} \) can be represented as:
\begin{align}
  u_{Bottle Beam} = u_{0,2}(r, \theta, z)\exp\left[i (1 - (2 - 0)) \pi\right] + u_{0,0}(r, \theta, z) 
\end{align}
orthogonally polarized waves do not interfere with each other, instead, the E fields of both beams add together and the corrected vectorial beam can be expressed as a Jones vector given by: 
\begin{align}
\mathbf{J}(r, \phi, z) = 
\begin{bmatrix}
\text{u}_{0,2}(r, \phi, z) + e^{i(1 - (2 - 1))\pi} \times \text{u}_{0,0}(r, \phi, z, ) \\
C \times \text{u}_{0,1}(r, \phi, z)
\end{bmatrix}
\end{align}

Where: \textit{C} is the scaling factor. The first component of the Jones vector represents the horizontally polarized Optical Bottle Beam. The second component of the Jones vector represents the correcting LG Beam with vertical polarization. The resulting local polarization is dependent on the combination of each of the local electric field magnitudes, directions, and phase offsets between the constituent beams. However, because the intensity distribution and the local polarization are both linked to the phase, only one of these two aspects can be optimized.

  Following Arlt, in the current paper, we have discussed the formation of optical bottles within an LG mode set. It is notable however that when such modes are highly focused the Jones vector expressed above must be modified to obtain the correct polarization and amplitude distribution.
  
  To understand the nodal surfaces in the beam a numerical simulation was done using Matlab. The bottle beam was represented as a complex vector wave. The beam was lensed by differential rotation of the electric field vectors and each point in the vector field is propagated and evaluated at successive planes in the focal volume. \cite{torok_electromagnetic_1997,munro_calculation_2007}. In Fig.\ref{Fig: Thoeretical intensity distribution of LG beams}(e) the focal volume is observed, and shells of uniform intensity show two bright nodes on the axis, before and after the dark focus while in the focal plane, the intensity forms a toroidal ring. Nodal surfaces, shown in red, follow a circular phase continuity through the focus creating conical surfaces with reduced intensity. 

\subsection{Beam Shaping }
\begin{figure}[ht!]
  \centering\includegraphics[width=9cm]{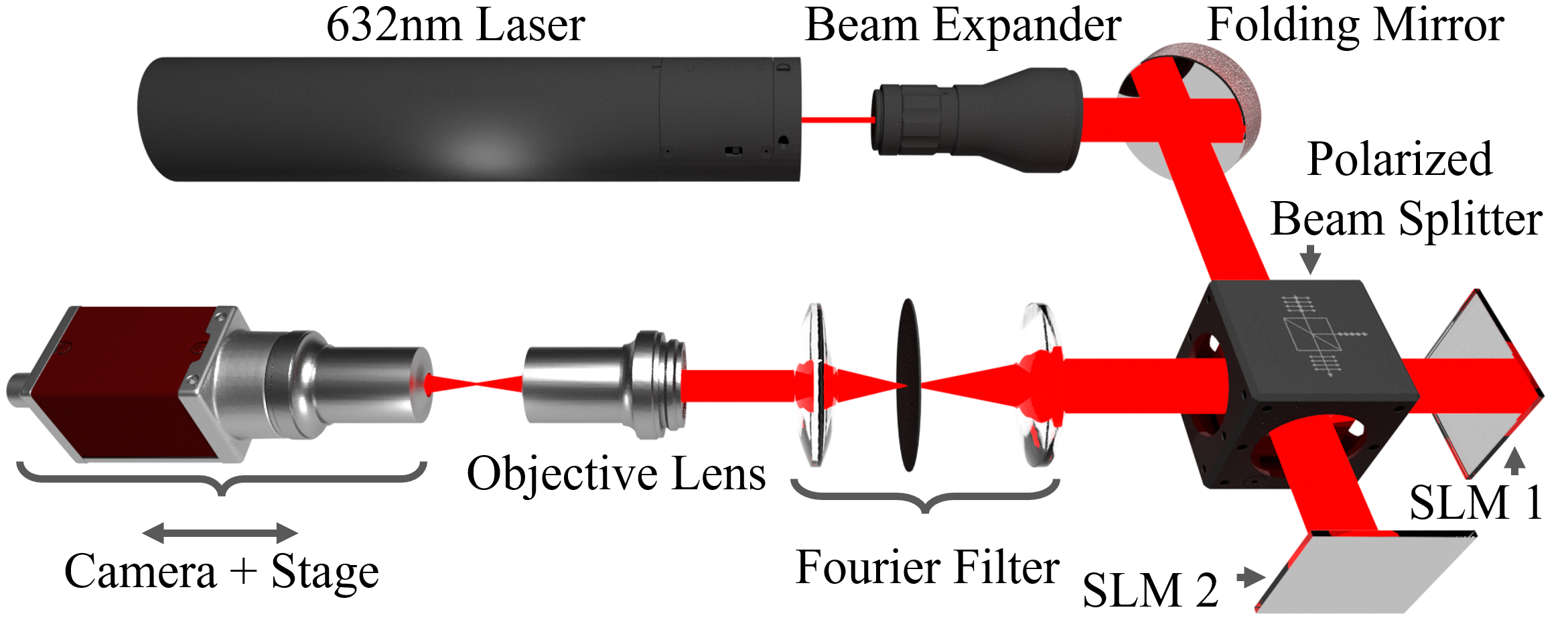}
  \caption{Schematic diagram of the experimental setup. An unpolarized HeNe laser is collimated and expanded before splitting polarizations by a polarized beam splitter (PBS) onto two twin spatial light modulators. The modulated beams are recombined by the PBS, filtered, and focused by a 0.25NA objective lens. The beam is imaged by a 1.25NA objective and scanned through the propagation axis of the beam.}
  \label{Fig: Schematic diagram of the experimental setup.}
\end{figure}
  
\begin{figure}[ht!]
  \centering\includegraphics[width=9cm]{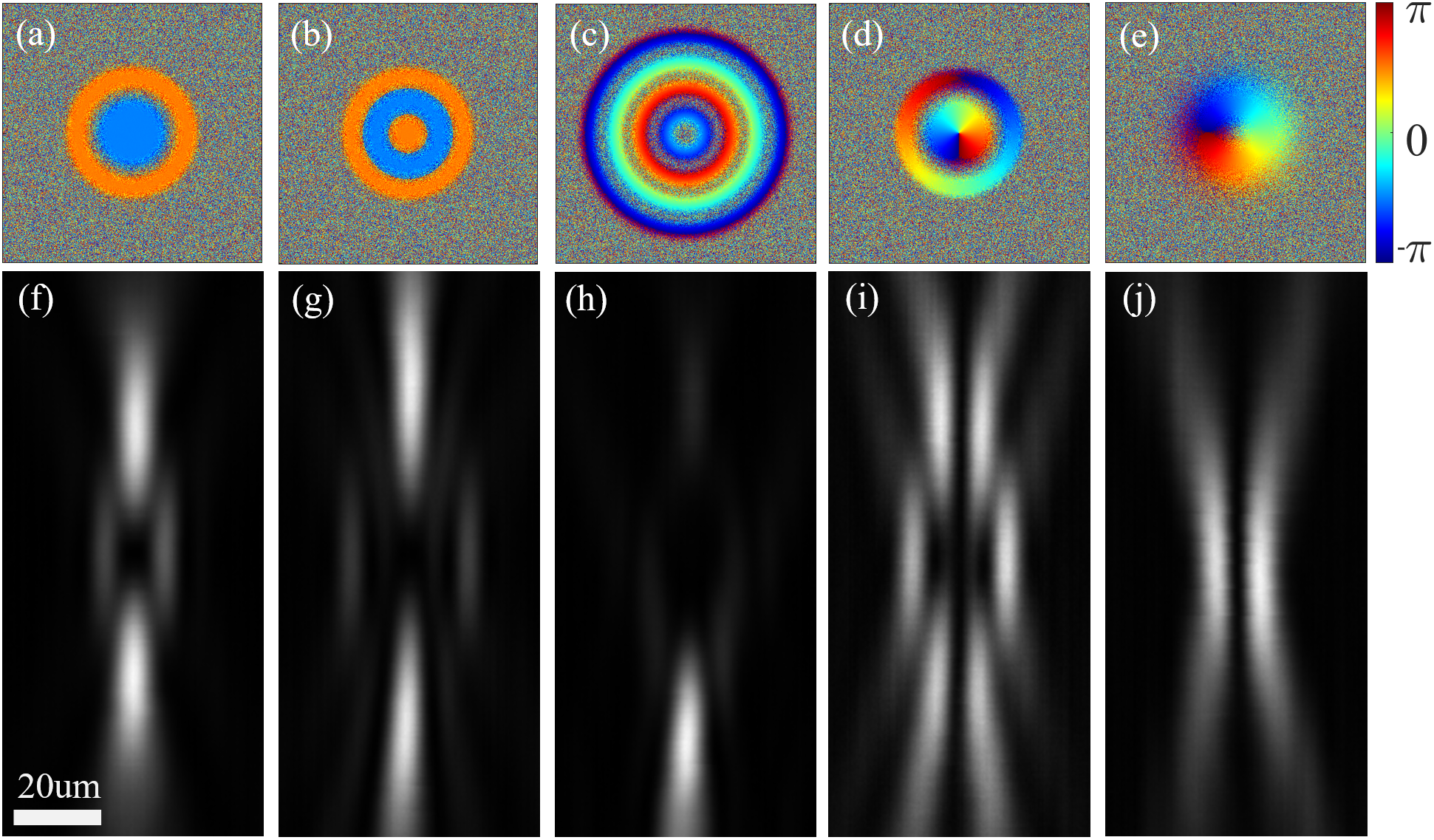}
  \caption{Phase and experimental cross sections of various bottle beams. (a-e) Beam phases of single beams using a single SLM. (f-j) respective beam cross sections captured by stitching together beam profiles scanned along the beam propagation axis.}
  \label{Fig: Phase and experimental cross sections of various bottle beams.}
  \end{figure}
    
  The optical bottle beams are experimentally created by imprinting calculated phases on plane wave Gaussian beams using spatial light modulators before propagating through an objective. We test a select few holograms maintaining experimental parameters to compare each method and determine the best complimentary base structures. Each beam phase and resulting field amplitude is shown in Fig.\ref{Fig: Phase and experimental cross sections of various bottle beams.} 
  
  The phase-shifted beams in Fig.\ref{Fig: Phase and experimental cross sections of various bottle beams.} (a,f) and (b,g) utilize circular discontinuities, with a $\pi$ phase shift, propagating conically through the focus to create a caged dark region. This method, seen in work by Arlt \cite{arlt_generation_2000}, produces the potential well with the smallest radius. 
  
  The axicon beam, produced similarly to McGloin \cite{mcgloin_three-dimensional_2003}, Fig.\ref{Fig: Phase and experimental cross sections of various bottle beams.} (c, h), uses a radially modulated conical phase inclination. This phase results in a linear combination of intersecting amplitude modulated conical waves to form the bottle, which in includes an added acceleration due to the combination of the converging field created by the axicon phase combined with the focusing objective. The final result more closely resembles an airy beam trap created by Chremos \cite{chremmos_fourier-space_2011}. 
  
  The LG beams seen in Fig.\ref{Fig: Phase and experimental cross sections of various bottle beams.} (d, i) and (e, j) are created by imprinting a spiral phase on the Gaussian beam with LG winding number l = 1. The particular OAM state is chosen to complement the other trap shapes. The vortex bottle beam trap, proposed by Alpmann \cite{alpmann_holographic_2012}, Fig.\ref{Fig: Phase and experimental cross sections of various bottle beams.} (d, i), is created by the addition of OAM to the previous phase shifted beams. This results in a similarly lobed structure with an additional axial discontinuity. 

\section{Properties and Behavior of Enhanced Optical Bottle Beams}

\subsection{Orthogonally Polarized Beam Combination}
\begin{figure}[ht!]
  \centering\includegraphics[width=13cm]{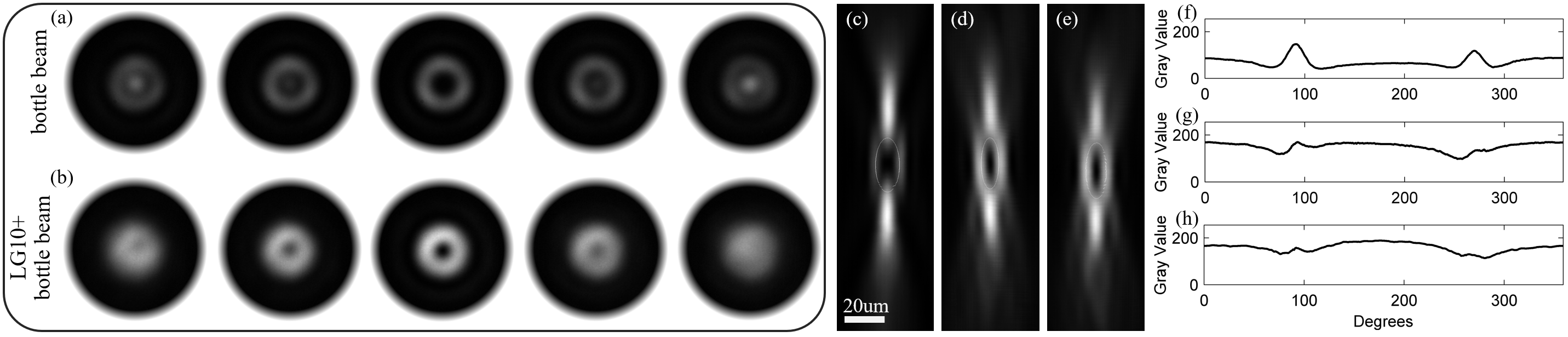}
  \caption{Beam profiles and bottle uniformity plots of multi-beam bottles. (a,b) experimental beam profiles of a bottle beam and obscured bottle beam at key points in the potential well. (c-e) experimental cross sections of orthogonally polarized beam combinations: (c) is the traditional bottle beam. (d,e) show the combination of traditional bottle beam combined with an orthogonal \( LG_{1,0} \) Beam , where in (e) the bottle beam is held at 75\% brightness. (f-h) Plot intensity vs. angle (originating east) of beams (c-e), respectively, displaying potential well uniformity.}
  \label{Fig: Beam profiles and bottle uniformity plots of multi-beam bottles.}
\end{figure}

  We utilize multiple coaxial non-interfering beams at the same wavelength to obscure the nodal surfaces seen in the single optical bottle beam geometry. Each of the traditional bottle beam shapes have nodal surfaces which can result in a "leaky" trap. This is unavoidable with the current methods of single beam shaping due to the usage of discontinuities to form the potential well. Using two beams at orthogonal polarization it is possible to overlap intensity patterns without creating additional nodal surfaces. Thus, finding two beams with complementary intensity profiles to obscure nodes and create a more uniform bottle beam solves this problem.
  As noted previously, the uniform bottle with the tightest dark region was created by combining Arlt and Padgett's bottle beam with an \( LG_{1,0} \) beam as seen in Fig.\ref{Fig: Beam profiles and bottle uniformity plots of multi-beam bottles.} (c). Other beam combinations show improvements shown in suplimentary data, however, higher-order nodal surface geometries create a more complex structure to complement and uniformly obscure. Illustrated by Fig.\ref{Fig: Thoeretical intensity distribution of LG beams}, \( LG_{1,0} \) and traditional bottle beams form naturally complementary shapes. Differential tuning of brightness, seen in Fig.\ref{Fig: Beam profiles and bottle uniformity plots of multi-beam bottles.}(e), and also beam waist may be employed to further enhance uniformity. 
  
\begin{figure}[ht!]
  \centering\includegraphics[width=7cm]{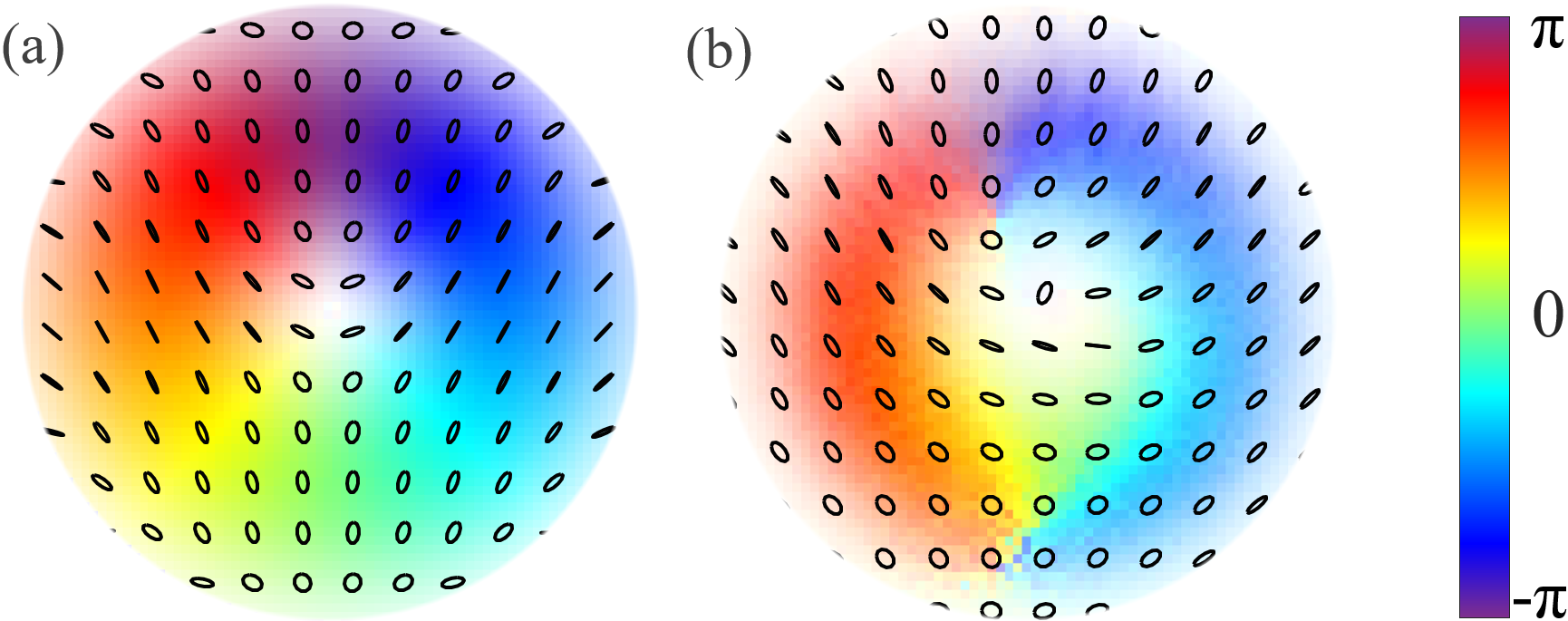}
  \caption{Polarization textures of vector simulated (a) and experimental (b) obscured bottle beam polarization in the focal plane, where the color bar represents local angle of polarization. In both cases, a horizontally polarized \( LG_{1,0} \) beam is combined with a vertically polarized bottle beam. Experimental polarization is retrieved through Stokes vector analysis from multiple simultaneous linearly polarized images.}
  \label{Fig: Polarization characterization of obscured bottle beam.}
\end{figure}

  Since spin angular momentum plays a role in many light-particle interactions, which by current methods is only controlled globally, to complete the characterization of the beam we also investigate the local polarization states of these beams. The resultant local polarization from the orthogonal combination of an \( LG_{1,0} \) and traditional bottle beam is asymmetrical. The helical phase delay, when combined with the radially symmetrical phase of the bottle beam leads to rotations of polarization angle and polarization ellipticity about the central axis of the beam.
  
\section{Discussion}
\subsection{Applications of High-Quality Optical Bottle Beams}

  Trap stiffness and stability are all important aspects of optical trapping at any scale. Optical traps, being three-dimensional potential wells, depend on the trap's spherical uniformity. By obscuring the nodal surfaces, the dips along the edges of the potential well are filled in. This will translate to an improvement in restoration force as particles are perturbed along these nodal surfaces.
  
  It is notable however that despite the symmetrical intensity in the obscured bottle beams. The vectorial nature of the beams lead to the presence of both spin and orbital angular momentum in the bright region surrounding the core. This can be seen in Fig. \ref{Fig: Polarization characterization of obscured bottle beam.}. It is outside the scope of this paper to analyze this in great detail, however, the presence of such momentum may impact the rotational torques on trapped particles.
  
  As can be noted from Fig.\ref{Fig: Thoeretical intensity distribution of LG beams}, it is remarkable how well an optimally shaped LG mode can fill in the dark areas around the central dark core of the bottle beam. It has been noted, several other bottle beam types exist with less trivial complementary beams which have not been considered in detail in the current paper. 
  
  With the orthogonal combination technique, differential tuning in three dimensions is possible in multiple ways. By attenuating the two beams separately before combination, the ratio of on-axis to lateral intensity may be adjusted. It is also possible to adjust the relative beam waist sizes by adjusting one beam waste in relation to the other, which can easily be executed using an SLM. Therefore, in a practical sense, when choosing complimentary beams, it is not completely necessary to begin with the perfect compliment. 
  
  The improvements to uniformity and tunability provided by this method should prove valuable to fields trapping atoms and other ultra-fine particles, that typically have trapping time scales on the order of a few seconds \cite{sun_influence_2020}. The obscured bottles should also show improvement in trapping reliability of particles with diameters much smaller than the beam waist. 
  
  Due to the experimental necessities of the current paper, we have not sought to shape the \(E_z\) component of the field except as a consequence of normal focusing. Doing so, however, would yield significantly more flexibility in the creation of optimized bottle beams, but would increase experimental complexity significantly. We also note that replacing the LG modes basis with an HG mode basis set leads to beams with a transverse angular momentum. This has been discussed more broadly in \cite{preece_rotational_2017}

\subsection{Bottle beams as quasi-particles.}
  Recently there has been significant interest in optical quasi-particles such as optical Skyrmions\cite{shen2022,mcwilliam2023topological}. Although the focus of the current work is optimizing bottle beams. We note that the current beam could be regarded in some senses as a quasi-particle. Such particles are conventionally difficult to make in the paraxial regime where the field vectors are exclusively transverse to the propagation direction. In the highly focused case, a longitudinal field component is introduced in the field structure enabling a 3D electromagnetic vector necessary for particle-like polarization topologies\cite{Du2019,gutierrez2021}.
  
  Experimental and theoretical data taken in the focal plane, seen in Fig. \ref{Fig: Polarization characterization of obscured bottle beam.} reveals polarization identical to a bimeron with skyrm number 1. We believe that our paper represents the first instance of an experimentally reproduced dark core quasi-particle. 


\section{Conclusion}

  High-quality optical bottle beams play a vital role in advancing research and technology. Using orthogonally polarized beams to combine complimentary beam shapes produces definitive improvements to field amplitude structure. These improvements to beam quality and the methods used to create them provide valuable tools for future cutting edge advancements.
  
\section{Acknowledgements:}

\begin{backmatter}

\bmsection{Funding}

We would like to acknowledge funding by the Air Force Office of Scientific Research under award number FA9550-17-1-0193 and the Beckman Laser Institute Foundation.

\bmsection{Disclosures}
none.
\end{backmatter}

\newpage
\section{References}

\bibliography{BottleBeams,skyrmion}

\newpage
\section{Supplementary:}
\subsection{Laguere Gaussian field amplitude:}
\begin{align}
{u}_{l,p}(r, \phi, z) = \sqrt{\frac{2}{\pi}} \frac{1}{w(z)} \left( \frac{\sqrt{2} r}{w(z)} \right)^{|l|} \exp\left(-\frac{r^2}{w(z)^2}\right) \exp(i l \phi) L_p^{|l|}\left(\frac{2 r^2}{w(z)^2}\right) \exp\left(-i \frac{k r^2}{2 R(z)}\right) \nonumber\\ \exp(ikz) \exp\left(-i (2p + |l| + 1) \arctan\left(\frac{z}{z_R}\right)\right)
\end{align}
Where:
- \( r \) and \( \phi \) are the radial and azimuthal coordinates.
- \( z \) is the axial coordinate.
- \( k \) is the wave number.
- \( w(z) \) is the beam waist at position \( z \).
- \( z_R \) is the Rayleigh range.
- \( l \) is the azimuthal index.
- \( p \) is the radial index.
- \( L_p^{|l|} \) is the associated Laguerre polynomial.
- \( R(z) \) is the radius of curvature of the wavefronts at position \( z \).

\subsection{The bottle beam field amplitude is the given by:}
\begin{align}
\text{BottleBeam}(r, \phi, z) = \text{u}_{0,2}(r, \phi, z) + \exp\left[i (1 - (2 - 0)) \pi\right] \text{u}_{1,0}(r, \phi, z)
\end{align}

\subsection{The corrected vectorial beam field amplitude can be expressed as a Jones vector given by:}
 
\begin{align}
\mathbf{J}(r, \phi, z) = 
\begin{bmatrix}
\text{u}_{0,2}(r, \phi, z) + e^{i(1 - (2 - 1))\pi} \times \text{u}_{0,0}(r, \phi, z, ) \\
C \times \text{u}_{0,1}(r, \phi, z)
\end{bmatrix}
\end{align}

Where:
\textit{C} is the scaling factor.

notes: The first component of the Jones vector represents the horizontally polarized Optical Bottle Beam. The second component of the Jones vector represents the correcting LG Beam with vertical polarization.

\subsection{Various tested beam combinations:}
\begin{figure}[ht!]
  \centering\includegraphics[width=13cm]{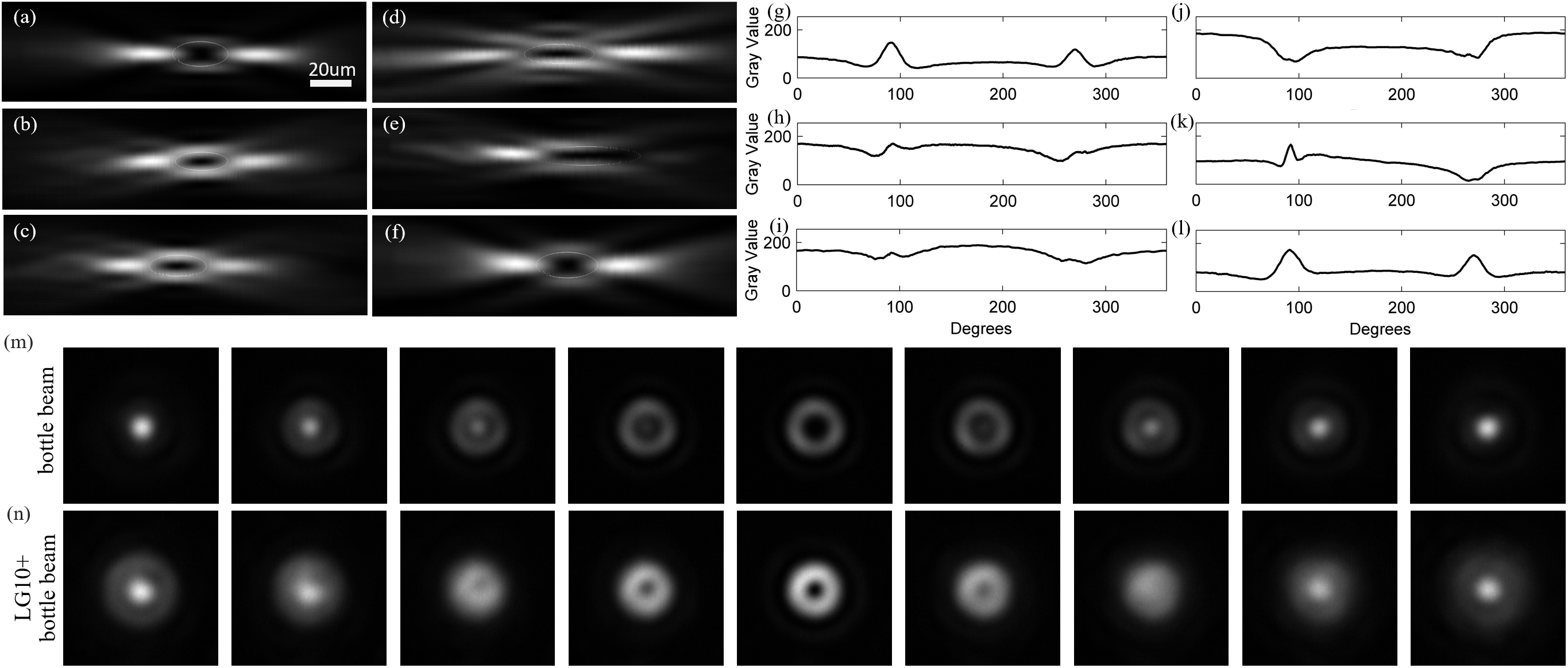}
  \caption{Beam profiles and bottle uniformity plots of multi-beam bottles. (a-f) experimental cross sections of orthogonally polarized beam combinations:(a) arlt bottle beam, (b) combination of Arlt bottle beam at 75\% power and \( LG_{1,0} \) beam, (c) combination of Arlt bottle beam and \( LG_{1,0} \) beam, (d) combination of second-order bottle beam and \( LG_{1,0} \) beam, (e) combination of lensed modulated Bessel beam and \( LG_{1,0} \) beam, (c) combination of Arlt bottle beam and second-order bottle. (g-l) represent potential well uniformity plots for beams (a-f) respectively. (m,n) show beam profiles at various key locations along the potential well for (a,b), respectively}
  \label{Fig: Beam profiles and bottle uniformity plots of multi-beam bottles supplemental.}
\end{figure}
\end{document}